\documentclass[pra,showkeys,twocolumn,showpacs]{revtex4-1}
\usepackage{amsmath}
\usepackage{amsfonts}
\usepackage{amssymb}
\usepackage{array}
\usepackage{graphicx}

\usepackage{xcolor}
\usepackage{hyperref}

\usepackage{color}
\usepackage{soul}

\newcommand{\Tr}{{\rm{Tr}}}

\begin{document}
\title{Super-quantum discord in ferromagnetic and antiferromagnetic materials}

\author{A. V. Fedorova} 
\affiliation{Institute of Problems of Chemical Physics of Russian Academy of Sciences, Acad. Semenov av. 1, Chernogolovka, Moscow
Region, Russia, 142432}

\author{Tim Byrnes}
\affiliation{New York University Shanghai, 1555 Century Ave, Pudong, Shanghai 200122, China}
\affiliation{State Key Laboratory of Precision Spectroscopy, School of Physical and Material Sciences,East China Normal University, Shanghai 200062, China}
\affiliation{NYU-ECNU Institute of Physics at NYU Shanghai, 3663 Zhongshan Road North, Shanghai 200062, China}
\affiliation{National Institute of Informatics, 2-1-2 Hitotsubashi, Chiyoda-ku, Tokyo 101-8430, Japan}
\affiliation{Department of Physics, New York University, New York, NY 10003, USA}

\author{A.~N.~Pyrkov} 
\email{pyrkov@icp.ac.ru}
\affiliation{Institute of Problems of Chemical
Physics of Russian Academy of Sciences, Acad. Semenov Av. 1, Chernogolovka, Moscow
Region, Russia, 142432}

\date{\today}

\begin{abstract}

Super-quantum discord is an extension of the quantum discord concept where weak measurements are made instead of projective measurements. We study the temperature behavior of the super-quantum discord for two real magnetic materials. We extract information about super-quantum discord from the magnetic susceptibility data for iron nitrosyl complexes $\mathrm{ Fe_2(SC_3 H_5N_2)_2(NO)_4}$ and binuclear Cu(II) acetate complex $\mathrm{[Cu_2 L(OAc)]\cdot 6 H_2O}$, where $\mathrm{ L}$ is a ligand that allows us to compare the super-quantum discord with the standard one. The dependence of the super-quantum discord on the parameter describing weak measurements is studied. Obtained difference between super-quantum and quantum discords confirms the detection of additional quantum correlations that are usually destroyed during projective measurements. The use of the approach allows to predict quantitatively in advance the advantages of use of weak measurements versus projective one for quantum technologies in real settings where quantum discord is used as resource. This can be relevant particularly in macroscopic quantum systems where weak measurements can be used to extract information about the system.

\end{abstract}


\maketitle

\section{Introduction}
\label{intro}

The presence of quantum correlations \cite{amico,streltsov17,horodecki09,bera18, bose, venuti,jetp07, sahling, pyrkov2013} allows for the possibility of superior performance of quantum devices over their classical counterparts \cite{nielsen,steane}. In particular, quantum correlations play a major role in quantum computation \cite{ekert,henderson,bennett}, quantum cryptography \cite{benbras,ekert91,gisin02, bennett92} and quantum metrology  \cite{layden,knill,shlyakov,orbru}, which demonstrate advantages in their performance over classical methods. Development of quantum technologies is also fundamental for realization of quantum communications lines and networks \cite{bose03, christandl, kuzzen, dorfeldzen, feldkuzzen}, as well as quantum simulation for development of novel materials.  
 Such materials should give opportunities to maintain and certify quantum correlations, and also contribute to the creation of materials for novel quantum devices. In the early stages of the field of quantum technologies, it was believed that entanglement was the necessary resource for a quantum advantage,  and many different efforts were developed to calculate entanglement \cite{amico,streltsov17,horodecki09,bera18, C1,wootters2,yurNMR,ciliberti}. However, it now better appreciated that not all quantum correlations can be associated with entanglement, and other concepts to measure quantum correlation were introduced  \cite{modi,ollivier}. One of the most well-known quantifiers of quantum correlations is quantum discord which allows one to distinguish quantum correlations from classical correlations with the use of an optimization over all possible projective measurements \cite{ollivier,A1}. 

Projective measurements are a particular type of quantum measurement which cause the collapse the wave function. For local projective measurements defined on subsystems, the quantum correlations are destroyed between the subsystems \cite{darwinism}, and the 'global' projective measurements in a basis of entangled states, which can produce quantum correlations \cite{gonzalez, kimble, sangouard, briegel}, are very difficult to implement \cite{kulik2001,schmidt2003,nolleke2013}. These are however not the only type of quantum measurement that is possible.  To reduce the influence of measurement on a system, a measurement that creates only a partial collapse (destruction) of a quantum state can be produced \cite{korotkov1,aharonov}. This kind of measurement is called a weak measurement, and allows one to preserve more quantum information after the measurement \cite{oreshkov,braginsky}. A scheme for weak measurements was introduced  by Aharonov, Albert and Vaidman in 1998 \cite{aharonov}. First experiments were performed in 2006 \cite{korotkov1,korotkov2}, and demonstrated the possibility of measuring a quantum state without destroying it completely. It allows for more quantum correlations to be preserved in the system after the weak measurement than with projective measurements. 

Recently there have been much interest in  realizing weak measurements in condensed matter systems \cite{pfender19, cujia18, lu14}. However, from resource point of view, it is not clear how to estimate the advantages of weak measurements quantitatively for quantum technologies. Furthermore, it is not clear whether is it possible to predict quantitatively advantages of weak measurements from a resource point of view for some real materials before realization the weak measurement?

In 2014, Singh and Pati generalized the concept of quantum discord in order to estimate the magnitude of quantum correlations during weak measurements \cite{singh}. They called the measure ``super-quantum discord'' and showed that it can be used for calculating quantum correlations after a sequence of weak measurements.  Super-quantum discord differs from the quantum discord --- it exceeds the quantum discord and becomes zero only for factorized states \cite{boli}, where the density matrix can be represented as a tensor product of the corresponding subsystems density matrices. Recently, the dependence of the quantum discord on temperature in antiferromagnetic copper nitrate $\mathrm{Cu(NO_3)_2\cdot 2.5H_2 O}$ and iron nitrosyl complexes was investigated \cite{aldoshin,spinklaster,ald14}.  Here we quantitatively investigate the difference between standard discord and super-quantum discord for two real magnetic materials with the use of standard data on magnetic susceptibility. We investigate the dependence of the super-quantum discord on temperature and the parameters describing weak measurements for antiferromagnetic iron nitrosyl complexes $\mathrm{Fe_2(SC_3H_5N_2)_2(NO)_4}$ \cite{sanina} and for binuclear ferromagnetic copper acetate complex $\mathrm{[Cu_2L(OAc)]\cdot 6H_2O}$ in order to show the boost in quantum correlations by using a weak measurement. For these two examples of materials it is shown that for one of them the use of weak measurements presents an excellent improvement from a resource point of view, while for the other compound the improvement is rather minor. Thus this approach allows one to estimate in advance  the parameters of weak measurements needed for advantages in the context of quantum technologies where quantum discord is used as resource. It can provide perspectives in calculating quantum discord especially between macroscopic qubits \cite{Byrnes2015, pyrkov19, gross12, pyrkov2014, byrnes12, pyrkov12, kunkel2017spatially, fadel2017spatial} where weak measurements can also be used to effectively characterize the system \cite{IloOkeke2014}.

\section{Discord and Super-discord}

The mutual information of two subsystems is a well-known measure of correlations in a bipartite system \cite{henderson,ollivier,modi}. In quantum information theory, the quantum mutual information of a bipartite system is defined as follows \cite{henderson,ollivier}:
\begin{equation}\label{Iro}
I(\rho_{AB})=S(\rho_A)+S(\rho_B)-S(\rho_{AB}),
\end{equation}
where $\rho_A$ and $\rho_B$  are the reduced density matrices of A and B subsystems $\rho_{AB}$ is the system matrix, and $S(\rho)$ is the entropy.
In quantum information theory,  $I(\rho_{AB})$ is a measure of both classical and quantum correlations between subsystems. The separation of classical and quantum correlations contribution into  $I(\rho_{AB})$  is complicated by the fact that in the quantum system measurements carried out on one subsystem can affect the another. The problem of separating classical and quantum correlations in a bipartite system was solved in Refs. \cite{henderson} and \cite{ollivier}. For example, classical correlations can be defined as the maximum information about subsystem $A$, which can be extracted by performing a complete set of subsystem $B$ projective measurements.  Henderson and Vedral proposed to determine the classical correlations according to the formula \cite{henderson}:
\begin{equation}\label{C}
C(\rho_{AB})=\max_{\{\Pi_i\}}[S(\rho_A)-\sum_i{p_i S(\rho_A^i)}],
\end{equation}
where $\{\Pi_i\}$ is the set of all projection-valued measurements performed on subsystem $B$, $\rho_A^i$  is the subsystem $A$ reduced density matrix, while the result of measurement performed on subsystem $B$ is $i$:
\begin{equation}
p_i\rho_A^i=\Tr_B\{\Pi_i\rho_{AB}\Pi_i^+\}
\end{equation}
and $p_i=\Tr_{AB} \{\Pi_i\rho_{AB}\Pi_i^+\}$ is the probability of the result $i$.
Since the correlations in the system are determined by the mutual information (\ref{Iro}), the quantum discord, which determines the quantum correlations in the bipartite system, will be the difference between the mutual information (\ref{Iro}) and the classical correlations (\ref{C}) \cite{zurek}:	
\begin{equation}
\label{discord}
D=I(\rho_{AB})-C(\rho_{AB})	. 
\end{equation}

Now let us consider the case that weak measurements are made.  The projective operators must be replaced by the following \cite{oreshkov}:
\begin{eqnarray}\nonumber
P(x)=\sqrt{\frac{1-\tanh(x)}{2}}\Pi_++\sqrt{\frac{1+\tanh(x)}{2}}\Pi_-,\\
P(-x)=\sqrt{\frac{1+\tanh(x)}{2}}\Pi_++\sqrt{\frac{1-\tanh(x)}{2}}\Pi_-,\nonumber
\end{eqnarray}
where $x$ is the measurement strength parameter, and $\Pi_\pm $ are two orthogonal projectors with $\Pi_++\Pi_-=1$. In the limit when $x\rightarrow\infty$ , we get projective measurements. For instance in NMR, where the weak measurements are realized via periodical coupling a nuclear spin to electronic spin, the measurement strength parameter is the interaction strength between the nuclear and electronic spins controlled by a dynamical decoupling sequence applied to the meter spin \cite{taminiau, boss}. In this case, it is possible to tune smoothly the measurement strength, or turn it off completely, by varying the interaction time \cite{pfender19, cujia18}.  The weak measurement operators satisfy the following properties:
\begin{enumerate}
	\item $P^+(x)P(x)+P^+(-x)P(-x)=1$.
	\item $P(0)=\frac{I}{\sqrt{2}}$.
	\item $\lim\limits_{x\rightarrow\infty}P(-x)=\Pi_+$; $\lim\limits_{x\rightarrow\infty}P(x)=\Pi_-$.	
\end{enumerate}
The measure of quantum correlations for weak measurements (super-quantum discord) is similar to the standard expression for quantum discord with projective measurements. For the case weak measurements Eq. (\ref{C})  is then replaced by
\begin{equation}
C_w(\rho_{AB})=\max_{\{\Pi_i\}} [S(\rho_A)-\sum\limits_{y=+x,-x}p(y)S(\rho_{A|P^B_{(y)}})],
\end{equation}
where
\begin{multline}
\rho_{A|P^B_{(y)}}=\frac{\Tr_A[(I\bigotimes P^B(y))\rho_{AB}(I\bigotimes P^B(y))]}{\Tr_{AB}[(I\bigotimes P^B(y))\rho_{AB}(I\bigotimes P^B(y))]}
 \\ =\frac{\Tr_A[(I\bigotimes P^B (y))\rho_{AB}(I\bigotimes P^B (y))]}{p(y)},
\end{multline}
where
\begin{equation}\nonumber
p(\pm x)=\Tr_{AB}[(I\bigotimes P^B(\pm x))\rho_{AB}(I\bigotimes P^B (\pm x))] . 
\end{equation}
Here,  $p(\pm x)$ is the probability of a state after a weak measurement. Thus, the super-quantum discord can be represented similarly to the normal discord (\ref{discord}):
\begin{equation}
\label{superdiscord}
D_w=I(\rho_{AB})-C_w(\rho_{AB}) .
\end{equation}

\section{Comparison of standard and super-quantum discords for real magnetic complexes}

We study the temperature dependence of the super-quantum discord from 5 to 300 K for antiferromagnetic iron nitrosyl complexes $\mathrm{Fe_2(SC_3H_5N_2)_2(NO)_4}$\cite{aldoshin} and for binuclear ferromagnetic copper acetate complex $\mathrm{[Cu_2L(OAc)]\cdot 6H_2O}$, where $\mathrm{H_3L=}$ 2-(2-hydroxyphenyl)-1,3-bis[4-(2-hydroxyphenyl)-3-azabut-3-enyl]-1,3-imidazolidine \cite{spinklaster,Fondo}.
In these materials the exchange interaction can be described in the frame of Heisenberg model of the magnetic dimer \cite{aldoshin, Fondo}:
$$
H=-\frac{1}{2}J\sigma_1\sigma_2,
$$
where $J$  is the exchange constant,  $\sigma_1=\sigma\bigotimes I$  and $\sigma_2=I\bigotimes \sigma$ . $I$ is the unit matrix, and $\sigma=(\sigma_x,\sigma_y,\sigma_z)$  is the Pauli Matrix vector
\begin{equation}
\sigma_x=\begin{pmatrix}
0 & 1\\
1 & 0
\end{pmatrix},\qquad 
\sigma_y=\begin{pmatrix}
0 & -i\\
i & 0
\end{pmatrix},\qquad
\sigma_z=\begin{pmatrix}
1 &0 \\0 & -1
\end{pmatrix},
\end{equation}

In the thermal equilibrium state, the density matrix of the system is given by:
$$
\rho=\frac{1}{Z}\exp\left(-\frac{H}{k_B T}\right),
$$
where $Z=\Tr\left[\exp\left(-
H/{k_BT}\right)\right]$ is the partition function. Since
$$
\sigma_1\sigma_2=
\begin{pmatrix}
1&0&0&0\\
0&-1&2&0\\
0&2&-1&0\\
0&0&0&1
\end{pmatrix},
$$
then the density matrix of the Heisenberg dimer is given by:
\begin{equation}\label{dm}
\rho(T) = \frac{1}{Z}
\begin{pmatrix}
e^{L}&0&0&0\\
0&e^{-L}\cosh(2L)&e^{-L}\sinh(2L)&0\\
0&e^{-L}\sinh(2L)&e^{-L}\cosh(2L)&0\\
0&0&0&e^{L}
\end{pmatrix}
\end{equation}
with $Z= 3e^{L}+e^{-3L}$ and $L=J/2k_BT$. One can easily verify using simple algebra that this is equivalent to
\begin{equation}\label{dm}
\rho(T) = \frac{1}{4}(1+G\sigma_1 \sigma_2)=\frac{1}{4}
\begin{pmatrix}
1+G&0&0&0\\
0&1-G&2G&0\\
0&2G&1-G&0\\
0&0&0&1+G
\end{pmatrix} 
\end{equation}
where
 \begin{equation}\label{corfun}
 G(T)=\frac{4}{3+\exp(-2J/k_BT)}-1 .
 \end{equation}
Furthermore, the $G$ represents the spin correlation functions that can be check by direct calculations
 $G=\langle \sigma_1^x \sigma_2^x \rangle=\langle \sigma_1^y \sigma_2^y \rangle=\langle \sigma_1^z \sigma_2^z \rangle$, where angle brackets denote statistical averaging.

The magnetic susceptibility in this case can be determined by the Bleaney-Bowers equation \cite{aldoshin}
\begin{equation}\label{blinibowers}
\chi(T)=\frac{2N_A g^2\mu_B^2}{k_B T(3+\exp(-2J/k_B T))}=\frac{N_A g^2 \mu_B^2}{2k_B T}(1+G(T)),
\end{equation}
where $N_A$ is the Avogadro constant, $k_B$ is the Boltzman constant, $T$ is the temperature, $g$ is the g-factor, if the measurement was provided on a single crystals. If the measurement was provided on a polycrystalline sample
$$
g^2=\frac{1}{3}(g_x^2+g_y^2+g_z^2).
$$
 $g_x,g_y,g_z$ are the component of a g-factor. 
From equations (\ref{blinibowers}) and (\ref{corfun}) the spin correlation function can be expressed in terms of the magnetic susceptibility (\ref{blinibowers})
$$
G(T)=\frac{2k_BT\chi(T)}{N_A g^2 \mu_B^2}-1=\frac{\chi(T)}{\chi^{curie}}-1,
$$
where $\chi^{curie}$ is the magnetization of the dimer determined by the Curie law. This formula allows the further using of experimental data for magnetic suspectibility for the calculations of the super-quantum discord \cite{bell}.

\begin{figure}[t]
\includegraphics[width=\columnwidth]{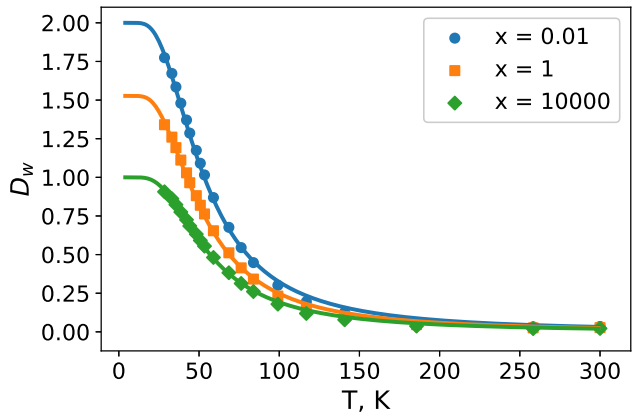}
\caption{The temperature dependence of a super-quantum discord in iron nitrosyl complexes for various values of the measurement strength parameter. Dots represent the super-quantum discord value, obtained from experimental data for magnetic susceptibility.}
\label{fig:1}       
\end{figure}

The density matrix (\ref{dm}) has a form of an $ X $-state\cite{xstates, bell}. This simplifies the computation of the standard quantum discord (\ref{discord}) and super-quantum discord (\ref{superdiscord})  significantly. In this case, equations in which there is no minimization over all possible complete sets of projective measurements can be used. As a result, the quantum discord for the state (\ref{dm}) is given by \cite{bell}.
\begin{equation} 
\label{tqd}
D=\frac{1+G}{4}\log(1+G)-\frac{1-G}{2}\log(1-G)
+\frac{1-3G}{4}\log(1-3G),
\end{equation}
and the super-quantum discord is given by \cite{bell}:
\begin{multline}
\label{sqd}
D_w= 1 + \frac{1-3G}{4} \log(\frac{1-3G}{4})\\
+\frac{3(1+G)}{4} \log(\frac{1+G}{4})
-\frac{1-G\tanh{x}}{2}\log(\frac{1-G\tanh{x}}{2})\\
-\frac{1+G\tanh x}{2}\log(\frac{1+G\tanh x}{2})
\end{multline}
The super-quantum discord (\ref{sqd}) converges to the normal discord (\ref{tqd}) when $x\rightarrow\infty$.

The experimental data on the magnetic susceptibility for iron nitrosyl complexes $\mathrm{ Fe_2(SC_3 H_5N_2)_2(NO)_4}$
was obtained from Ref. \cite{sanina}. For a more careful analysis, the subtraction of contribution of the impurity was done in Ref. \cite{aldoshin}. The fit to the Bleany-Bowers equation was obtained with parameters $J/k_B=-68$~K, $g=2$. Figure \ref{fig:1} shows that super-quantum discord greatly exceeds quantum correlations determined by the normal discord starting from liquid nitrogen temperatures. Thus, the antiferromagnetic iron nitrosyl complex is of particular interest from resource point of view because the super-quantum discord in this material reaches a value of 2 at low temperatures, which means it reaches the magnitude of mutual information.

\begin{figure}[t]
\includegraphics[width=\columnwidth]{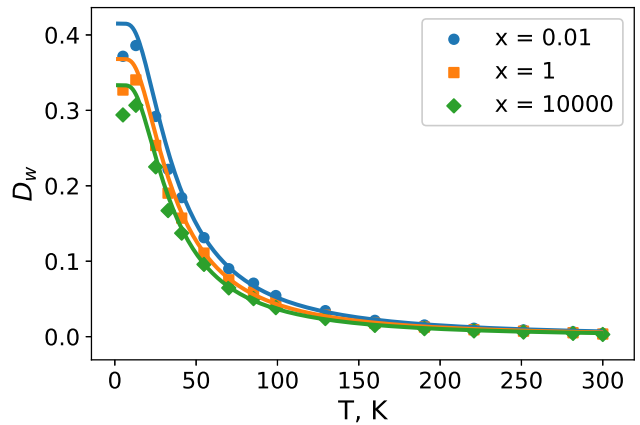}
\caption{The temperature dependence of a super-quantum discord in a binuclear copper acetate complex for various values of the measurement strength parameter. Dots represent the super-quantum discord value, obtained from experimental data for magnetic susceptibility.}
\label{fig:2}       
\end{figure}

In order to calculate super-quantum discord for the binuclear copper(II) acetate complex $\mathrm{[Cu_2L(OAc)]\cdot 6H_2O}$, where $\mathrm{H_3L=}$ 2-(2-hydroxyphenyl)-1,3-bis[4-(2-hydroxyphenyl)-3-azabut-3-enyl]-1,3-imidazolidine we use the magnetic susceptibility data in Ref. \cite{Fondo} for temperatures from 5 to 300 K which can be described with the Bleaney-Bowers equation derived  from  the  Heisenberg spin Hamiltonian taking  into  account  the  crystal  structure  of  this  compound \cite{Fondo}. The best fit was obtained with the parameter $2J=49.2 cm^{-1}$ ($J/k=35.4$ K) and $g=2.13$ \cite{Fondo}. Fig.\ref{fig:2} shows the temperature dependence of the super-quantum and standard discords obtained for the different values of parameter $x$ describing weak measurements. As before for the antiferromagnetic iron nitrosyl complexes, at low temperatures super-quantum discord exceeds quantum correlations as calculated by the standard discord but the difference is not as large as for the antiferromagnetic iron nitrosyl complexes. Thus this compound can be used as an example where using a weak measurement approach is less important. Also we can see that the experimental points are slightly lower than the theoretical predictions for the low temperatures. In the work of Ref. \cite{Fondo}, it was shown that attempts to evaluate the possible interdimer  interactions led  to  extremely low values without any significant improvement of the fitting and is rather difficult to take into account the influence of weak interdimer couplings. Thus we suggest that the influence of weak interdimer couplings is the reason for the mismatch between the experimental points and theoretical predictions.

\section{Conclusions}

In this work we calculated the temperature dependence of super-quantum discord as a function of the measurement strength parameter for iron nitrosyl complex and binuclear copper acetate complex. Super-quantum discord allows to reveal more quantum correlations than standard quantum discord, and for antiferromagnetic iron nitrosyl complexes at small measurement strength parameter values it reaches the magnitude of mutual information. On the other hand, the use of weak measurements is not as beneficial for the binuclear copper acetate complex. Thus it is shown that it is possible to quantitatively estimate advantages of weak measurement for real materials from a resource point of view in advance using standard magnetic data. Super quantum discord can be applied for determination the effectiveness of developed materials, with the aim of their further using for implementation of quantum technologies.

Authors thank E.B. Feldman and M.A. Yurishev for stimulating discussions.

The work is supported by the RFBR-NSFC collaborative program (Grant No. 18-57-53007). T. B. is supported by the Shanghai Research Challenge Fund; New York University Global Seed Grants for Collaborative Research; National Natural Science Foundation of China (61571301,D1210036A); the NSFC Research Fund for International Young Scientists (11650110425,11850410426); NYU-ECNU Institute of Physics at NYU Shanghai; the Science and Technology Commission of Shanghai Municipality (17ZR1443600); the China Science and Technology Exchange Center (NGA-16-001); and the NSFC-RFBR Collaborative grant (81811530112).

\end{document}